# Angiotensin II cyclic analogs as tools to investigate AT$_1$R biased signaling mechanisms

David St-Pierre, Jérôme Cabana, Brian J. Holleran, Élie Besserer-Offroy, Emanuel Escher,

Gaétan Guillemette, Pierre Lavigne, and Richard Leduc[#]

Department of Pharmacology-Physiology, Faculty of Medicine and Health Sciences,

Université de Sherbrooke, Sherbrooke, Quebec, Canada, J1H 5N4

**Running title: AT$_1$R biased signaling by cyclic Angiotensin II analogs**

[#]Corresponding author

Richard Leduc, Ph.D.
Department of Pharmacology-Physiology
Faculty of Medicine and Health Sciences
Université de Sherbrooke
Sherbrooke, Quebec
Canada, J1H 5N4
Tel.: (819) 821-8000 ext. 75413
Fax: (819) 564-5400
E-mail: Richard.Leduc@USherbrooke.ca

**ABSTRACT**

G protein coupled receptors (GPCRs) produce pleiotropic effects by their capacity to engage numerous signaling pathways once activated. Functional selectivity (also called biased signaling), where specific compounds can bring GPCRs to adopt conformations that enable selective receptor coupling to distinct signaling pathways, continues to be significantly investigated. However, an important but often overlooked aspect of functional selectivity is the capability of ligands such as angiotensin II (AngII) to adopt specific conformations that may preferentially bind to selective GPCRs structures. Understanding both receptor and ligand conformation is of the utmost importance for the design of new drugs targeting GPCRs. In this study, we examined the properties of AngII cyclic analogs to impart biased agonism on the angiotensin type 1 receptor ($AT_1R$). Positions 3 and 5 of AngII were substituted for cysteine and homocysteine residues ([Sar[1]Hcy[3,5]]AngII, [Sar[1]Cys[3]Hcy[5]]AngII and [Sar[1]Cys[3,5]]AngII) and the resulting analogs were evaluated for their capacity to activate the Gq/11, G12, Gi2, Gi3, Gz, ERK and β-arrestin (βarr) signaling pathways *via* $AT_1R$. Interestingly, [Sar[1]Hcy[3,5]]AngII exhibited  potency and full efficacy on all pathways tested with the exception of the Gq pathway. Molecular dynamic simulations showed that the energy barrier associated with the insertion of residue Phe[8] of AngII within the hydrophobic core of $AT_1R$, associated with Gq/11 activation, is increased with [Sar[1]Hcy[3,5]]AngII. These results suggest that constraining the movements of molecular determinants within a given ligand by introducing cyclic structures may lead to the generation of novel ligands providing more efficient biased agonism.

## GRAPHICAL ABSTRACT

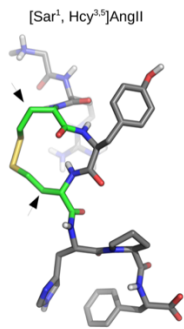

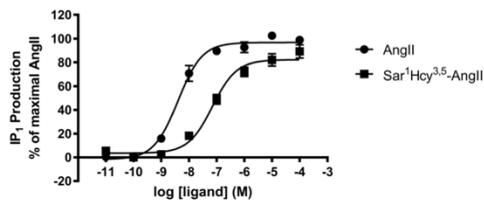

Cyclic AngII analog exhibited  potency and full efficacy on all pathways tested with the exception of the Gq pathway

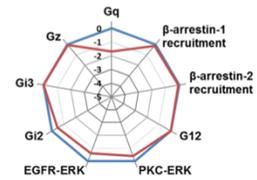

## 1. INTRODUCTION

The octapeptide hormone angiotensin II (AngII) is the active component of the renin-angiotensin system, responsible for controlling blood pressure and water retention via smooth muscle contraction and ion transport. It exerts a wide variety of physiological effects, including vascular contraction, aldosterone secretion, neuronal activation, and cardiovascular cell growth and proliferation. Virtually all the known physiological effects of AngII are produced through the activation of the $AT_1$ receptor ($AT_1R$), which belongs to the G protein-coupled receptor (GPCR) superfamily [1] and whose structure in complex with a selective $AT_1R$ antagonist has recently been elucidated [2].

The $AT_1R$ interacts with Gq/11 leading to the activation of phospholipase C (PLC), in turn leading to the formation of diacylglycerol (DAG) and inositol 1,4,5 triphosphate ($IP_3$). $IP_3$ binds to the $IP_3$ receptor on the endoplasmic reticulum, whereupon $Ca^{2+}$ is released into the cytosol. Together, $Ca^{2+}$ and DAG allow the activation of protein kinase C (PKC) [3]. Also, $AT_1R$ interacts with G12, thereby activating RhoA and ROCK, via RhoGEF regulation, leading to cytoskeleton reorganization [4]. Additionally, reports have demonstrated that $AT_1R$ interacts with Gi, thereby leading to an inhibition of cAMP production [5–7]. The $AT_1R$ can also activate the ERK1/2 kinase pathway mediated by PKC (G protein-dependent) or by EGFR transactivation, which is G protein-independent [8,9]. Following receptor activation, G protein-coupled receptor kinases (GRKs) phosphorylate the $AT_1R$, facilitating β-arrestin (βarr) recruitment and terminating G protein signaling [10]. βarrs are involved in the desensitization and internalization of GPCRs [11] but also serve as scaffolds for further GPCR signaling to the MAPK pathway. [12]

Biased signaling is the ability of a ligand to stabilize a receptor under a particular conformation that promotes activation of specific signaling pathways over others [13,14]. The therapeutic potential of functional selectivity is increasingly exploited for the design of new drugs since some signaling pathways produce beneficial effects while others can have harmful consequences. For example, activation of the Gq/11 pathway by AngII may cause adverse effects to the failing heart by increasing blood pressure, while βarr recruitment can produce a beneficial effect by promoting cardiomyocyte growth, thus improving heart performance [15,16]. The therapeutic potential of $AT_1R$ signaling via the G12 and Gi pathways has yet to be evaluated.

Based on extensive photolabeling experiments [17], we have demonstrated that the conformation of AngII is highly dynamic even when interacting with the $AT_1R$. We have also proposed an integrative model of this complex and unveiled structural and dynamical determinants that favor Gq/11 or βarr signaling. Of note, we have shown that Gq/11 signaling is promoted by the opening of a hydrophobic core just above the so-called major H-Bond network (MHN) or sodium binding site [18], notably by the insertion in this core of the C-terminal $Phe^8$ residue of AngII. In addition, the perturbation of the MHN was shown to modulate the signaling outcome. Our model of the dynamical AngII-$AT_1R$ complex and molecular dynamics simulations also suggest that the backbone of AngII can adopt multiple conformations.

We have recently shown that changes at positions 4 and 8 of AngII can lead to biased signaling of the $AT_1R$ [19]. To further pursue our understanding of the molecular and dynamical basis of functional selectivity, we asked how constraining the AngII backbone and thus limiting its dynamical behavior would impact the signaling pathways of the $AT_1R$. Therefore, we synthesized AngII analogs that were substituted at positions 3 and 5 with either cysteine or homocysteine (Hcy) and cyclised through the formation of a disulfide bond. Using this strategy, we synthesized [Sar[1]Hcy[3,5]]AngII, [Sar[1]Cys[3]Hcy[5]]AngII and [Sar[1]Cys[3,5]]AngII. This cyclisation scheme through the oxidation of the sulfhydryl group has the advantage of limiting the changes in physiochemical properties and bulkiness of the cycle. Note that previous work has shown that cyclisation through these side-chains of AngII led to compounds that can still bind and elicit Gq/11 activation [20,21]. However, the impact on the biased signaling has not been characterized. Here, we investigate the impact of restraining the conformation of AngII on the binding and functional selectivity of the $AT_1R$ by measuring the Gq, G12, Gi2, Gi3, Gz, ERK and βarr signaling pathways. We then performed molecular dynamics simulations to evaluate the impact of AngII cyclic analogs on the conformational landscape of the $AT_1R$.

## 2. MATERIALS AND METHODS

### 2.1 Materials

Culture media, trypsin, FBS, penicillin, and streptomycin were from Wisent (St-Bruno, Qc, Canada). OPTI-MEM was from Invitrogen Canada Inc. (Burlington, ON). Polyethyleneimine (PEI) was from Polysciences (Warrington, PA). Go6983 and PD168393 were from EMD Millipore (Missisauga, ON). $^{125}$I-AngII (specific radioactivity ~1000 Ci/mmol) was prepared with Iodo-GEN® (Perbio Science, Erembodegem, Belgium) as reported previously [22].

### 2.2 Peptide synthesis

Peptides were synthesized by manual solid-phase peptide synthesis using Fmoc-protected strategy on Wang resin (Fmoc-protected amino acids and resin were purchased from Novabiochem, Missisauga, ON). Peptides were cleaved with 95% TFA adding EDT and TLS as scavengers. The crude peptides were then cyclized in 2 M $(NH_4)_2CO_3$, pH 6.5, under constant agitation for 4 h at room temperature. Peptides were then purified on a preparative HPLC mounted with a $C_{18}$ column and using a 10–35% gradient of acetonitrile containing 0.05% TFA. Fractions were analysed on an analytical HPLC mounted with a $C_{18}$ column and using a 5–95% acetonitrile gradient containing 0.05% TFA. Pure fractions were pooled, lyophilized, and stored at -20 °C in a dry environment until further use. The pure peptides were characterized on UPLC-MS and showed purity >95%. Structure of the compounds and UPLC-MS spectra are available through Figshare at doi: 10.6084/m9.figshare.6108440.

### 2.3 Constructs

The cDNA clone for the human $AT_1R$ was kindly provided by Dr. Sylvain Meloche (University of Montréal). The $AT_1R$-GFP10 construct was built by inserting the GFP10 sequence at the C-terminus of the $AT_1R$, joined by a linker sequence (GSAGT) using the In-Fusion® PCR cloning system (Clontech Laboratories, Mountain View, CA) as recommended by the manufacturer. The RLucII-βarr1, RLucII-βarr2, Gα12-RLucII, Gαi2-RlucII, Gαi3-RlucII, Gz-RlucII, Gβ1, Gγ1-GFP10 and Gγ2-GFP10 constructs were kindly provided by Dr. Michel Bouvier (University of Montréal). All constructs were confirmed by automated DNA sequencing.

### 2.4 Cell culture and transfection

Human embryonic kidney 293 (HEK293) cells were maintained in DMEM medium supplemented with 10% FBS, 100 IU/ml penicillin, and 100 μg/ml streptomycin at 37°C in a humidified 5% $CO_2$ atmosphere. HEK293 cells stably expressing the $AT_1R$ were maintained in medium containing 0.5 mg/mL G418. For the βarr recruitment assays, HEK293 cells ($3 \times 10^6$ cells) were transiently transfected with 8700 ng of $AT_1R$-GFP10 and either 300 ng of RlucII-βarr1 or 300 ng of RlucII-βarr2 using linear polyethylenimine (1 mg/ml) (PEI:DNA ratio 4:1). For G12 activation assays, HEK293 cells ($3 \times 10^6$ cells) were transiently cotransfected with the following constructs: 3000 ng of $AT_1R$, 600 ng $G\alpha12$-RLucII, 3000 ng $G\gamma1$-GFP10 and 1800 ng $G\beta1$, using linear polyethylenimine (PEI:DNA ratio 4:1). For Gi2 and Gi3 activation assays, HEK293 cells ($3 \times 10^6$ cells) were transiently cotransfected with the following constructs: 3000 ng of $AT_1R$, 600 ng $G\alpha i2$-RLucII or $G\alpha i3$-RLucII, 3000 ng $G\gamma2$-GFP10 and 3000 ng $G\beta1$, using linear polyethylenimine (PEI:DNA ratio 3:1). For Gz activation assays, HEK293 cells ($3 \times 10^6$ cells) were transiently cotransfected with the following constructs: 3000 ng of $AT_1R$, 600 ng Gz-RLucII, 3000 ng $G\gamma1$-GFP10 and 3000 ng $G\beta1$, using linear polyethylenimine (PEI:DNA ratio 3:1).

*2.5 Binding Experiments*

HEK293 cells stably expressing the $AT_1R$ were washed once with PBS and submitted to one freeze-thaw cycle. These broken cells were then gently scraped into washing buffer (25 mM Tris-HCl, pH 7.4, 100 mM NaCl, 5 mM $MgCl_2$), centrifuged at $2500 \times g$ for 15 min at 4 °C, and resuspended in binding buffer (25 mM Tris-HCl, pH 7.4, 100 mM NaCl, 5 mM $MgCl_2$, 0.1% bovine serum albumin, 0.01% bacitracin). Dose displacement experiments were done by incubating broken cells (20–40 μg of protein) for 1 h at room temperature with 0.8 nM $^{125}$I-AngII as tracer and increasing concentrations of AngII or analogs. Bound radioactivity was separated from free ligand by filtration through GF/C filters presoaked for at least 3 hours in binding buffer. Receptor-bound radioactivity was evaluated by γ counting. Results are presented as means ± S.D. The $K_i$ values in the displacement studies were determined from the $IC_{50}$ values using the Cheng-Prusoff equation.

## 2.6 Measuring inositol-1 phosphate production

We used the IP-One assay (Cisbio Bioassays, Bedford, MA) to measure inositol 1-phosphate ($IP_1$) levels. Necessary dilutions of each analog were prepared in stimulation buffer (10 mM Hepes, 1 mM $CaCl_2$, 0.5 mM $MgCl_2$, 4.2 mM KCl, 146 mM NaCl, 5.5 mM glucose, 50 mM LiCl, pH 7.4). HEK293 cells stably expressing the $AT_1R$ were washed with PBS at room temperature, then trypsinized and distributed at 15 000 cells/well (7 µl) in a white 384-well plate in stimulation buffer. Cells were stimulated at 37°C for 30 min with increasing concentrations of AngII or analogues. Cells were then lysed with the lysis buffer containing 3 µl of $IP_1$ coupled to the d2 dye. After addition of 3 µl of anti-$IP_1$ cryptate terbium conjugate, cells were incubated for 1 h at room temperature under agitation. FRET signal was measured using a TECAN M1000 fluorescence plate reader (TECAN, Austria).

## 2.7 BRET-based biosensor assays

After 48 h post-transfection, cells were washed with PBS and resuspended in BRET buffer (10 mM Hepes, 1 mM $CaCl_2$, 0.5 mM $MgCl_2$, 4.2 mM KCl, 146 mM NaCl, 5.5 mM glucose, pH 7.4). For the βarr recruitment assays, the proximity of fusion protein RLucII-βarr to the reporter $AT_1R$-GFP10 was evaluated. Upon stimulation, RLucII-βarr was recruited to the $AT_1R$-GFP10 fusion protein, whereby the BRET signal was increased.

For the G12, Gi2, Gi3 and Gz activation assays, the biosensor measures the proximity of the fusion protein RLucII-Gα to GFP10-Gγ. Upon activation, both RLuc-Gα and GFP10-Gγ move away from each other, resulting in a decrease in the measured BRET. For all BRET assays, cells transfected with the appropriate constructs were stimulated with the indicated ligands in 96-well white plates (50 000 cells/well). Cells were stimulated for either 1 min (Gi2 and Gi3), 5 minutes (Gz) or 8 minutes (G12), and then coelentherazine 400A was added at a final concentration of 5 uM. All BRET signals were measured using a TECAN M1000 fluorescence plate reader. The BRET ratio was calculated as the GFP10 emission over luminescence emission. Ligand-promoted BRET ratio was calculated by subtracting the BRET ratio under basal conditions from the BRET ratio upon maximal stimulation. All data were expressed as a percentage of maximal AngII response.

*2.8 ERK1/2 activation assay*

ERK1/2 activation was measured using the ERK1/2 AlphaScreen Surefire kit (PerkinElmer, Waltham, MA). HEK293 cells stably expressing the $AT_1R$ were seeded into 96-well plates at a density of 125 000 cells/well. After 24 h, cells were starved for at least 16 h in phenol red-free media before stimulation. We previously showed that, in HEK293 cells stably expressing the $AT_1R$, the two main pathways leading to the activation of ERK1/2 by AngII were dependent on the activity of PKC and on the activity of EGFR. In the presence of both the EGFR tyrosine kinase inhibitor PD168393 and the PKC inhibitor Go6983 no significant ERK1/2 activity could be detected. Therefore, upon treatment with PD168393, the ERK response was PKC-dependent with a maximal response obtained after 2 min, while upon treatment with Go6983, the ERK response was EGFR-dependent with a maximal response obtained after 5 min [19]. Where specified, Go6983 (1 uM) or PD168393 (250 nM) were added 30 min before stimulation. For concentration-response experiments, cells were stimulated with increasing concentrations of indicated ligand. Stimulation of cells was terminated by the addition of lysis buffer to each well. The plate was then agitated at room temperature for 10 min and 4 μl of lysate was transferred to 384-well ProxiPlates (PerkinElmer, Waltham, MA) and 5 ul of the assay reaction mix was added to each well (reaction buffer : activation buffer : donor beads : acceptor beads = 120 : 40 : 1 : 1). The plate was then incubated in the dark at room temperature for 24 h under agitation and the signal was measured with an Enspire alpha reader (PerkinElmer, Waltham, MA) using standard AlphaScreen settings. All data were expressed as a percentage of maximal AngII-induced ERK1/2 phosphorylation.

*2.9 Data analysis*

Binding data were analyzed with Prism version 7.0 (GraphPad Software, San Diego CA), using a one-site binding hyperbola nonlinear regression analysis. Transduction ratios and bias factors were calculated based on the method of Kenakin [23], as described in detail by van der Westhuizen *et al* [24]. Transduction ratios [$\log(\tau/K_A)$] were first derived using the operational model equation in GraphPad Prism. The transduction ratio is an assessment of the effect (potency and efficacy) of a compound on receptor conformation and the subsequent ligand-receptor interaction with downstream effectors. In order to assess true ligand bias, system and observational bias which may be present owing to the different sensitivities of the assays used must be eliminated by comparing ligand activity at a given signaling pathway to that of a reference agonist. AngII, which yielded similar potencies and maximally activated all the pathways, was the reference compound. By subtracting $\log(\tau/K_A)$ of AngII to the $\log(\tau/K_A)$ value of each analog for a given pathway, a *within-pathway* comparison was first established, yielding $\Delta\log(\tau/K_A)$. Finally, *between-pathway* comparisons were achieved for a given ligand in the form of $\Delta\Delta\log(\tau/K_A)$ and the bias factor BF. $\Delta\Delta\log(\tau/K_A)$ were calculated by substracting $\Delta\log(\tau/K_A)$ values of one signaling pathway from the $\Delta\log(\tau/K_A)$ of the signaling pathway to which it is compared. BF values are the base 10 values of $\Delta\Delta\log(\tau/KA)$ and are the actual bias factors.

Statistical analyses of the $\Delta\Delta\log(\tau/K_A)$ values were performed with Prism version 7.0 using the Two-tailed unpaired student t-test. A value was considered statistically significant when $P < 0.05$.

### 2.10 Numbering of residues

In order to facilitate comparison of the same position of an amino acid across multiple GPCRs, residues in transmembrane domains (TMD) of the $AT_1R$ were given two numbering schemes. First, residues were numbered according to their positions in the $AT_1R$ primary structure. Second, residues were also indexed according to their position relative to the most conserved residue in each TMD in which they are located. By definition, the most conserved residue was assigned the position index "50" e.g. in TM6,

P255 is the most conserved residue and was designated P255[(6.50)], whereas the upstream residue was designated I254[(6.49)] and the downstream residue H256[(6.51)].

### 2.11 Molecular dynamic simulations

The GROMACS software suite [25,26] was used to prepare and run the simulations in a similar fashion to previous work [18]. The $AT_1R$ was inserted in a lipid bilayer consisting of 128 molecules of POPC using the *g_membed* tool in GROMACS. The membrane-receptor system was solvated with the SPC water model [27]. Sodium and chloride ions were added at random positions, replacing water molecules, to keep the net charge of the system at 0 and to have an approximate salt concentration of 150 mM. The ffg54a7 force field and the corresponding POPC parameters included in the GROMACS installation were used for the calculations [28,29]. A first equilibration phase was performed under NPT conditions for 1 ns while gradually heating the system for the first 500 ps to reach the desired temperature of 310 K. During this first phase, phosphate head group of the POPC molecules were restrained. This was followed by a second equilibration under NPT conditions for 100 ns with the pressure set at 1 bar, without the restraints on the POPC molecules. Such long equilibration was performed to allow proper equilibration of the lipids after embedding a protein in the membrane [30]. These equilibration times were reduced to 100 ps for the first phase and 1 ns for the second phase for all other simulations that were generated using the originally equilibrated system. The position of all heavy atoms of the receptor and ligand were restrained during equilibration. Unrestrained MD simulations were run in 5 fs steps for 1 μs of total MD simulation time, in the form of 10 simulations of 100 ns in length for each receptor. Using multiple shorter MD trajectories rather than a single long one prevents known problems that can occur with long trajectories in current force fields [31]. Random seed was used for velocity generation during the initial equilibration phase. Thus, all MD trajectories of a given system (e.g. AngII-AT$_1$R complex) share the same initial conformation and velocities at the start of the production MD runs. Divergence in the multiple trajectories is the result of the inherent imprecision of the calculations, which originate from the single-precision floating point format used as well as differences in the order of addition of force caused by dynamic load balancing ((a+b)+c $\neq$ a+(b+c) due to rounding-off). The simulations were run under periodic boundary conditions at constant temperature (310 K) and pressure (1 bar) using the Nose-Hoover thermostat [32,33] with $\tau$T = 2 ps and the Parrinello-Rahman barostat with $\tau$P = 5 ps, respectively. Simulation data were saved every 50 ps, for a total of 20001 frames per microsecond.

*2.12 Trajectory analysis*

MD trajectory outputs from GROMACS were converted to PDB files for visual inspection with PyMOL [34] and to compressed XTC trajectory files for other analyses. Distance between atoms or group of atoms were measured with the *g_dist* tool within GROMACS. One-dimensional probability distributions were calculated using the *g_analyze* tool. Two-dimensional probability density functions were calculated using *g_sham* with a grid of 50 x 50 bins and *nlevels* = 200.

# 3. RESULTS

## 3.1 Binding properties of cyclic AngII analogs

We evaluated the binding properties of three cyclic Ang II analogs. [Sar$^1$Hcy$^{3,5}$]AngII showed binding affinity in the low nanomolar range ($K_d$ = 2.7 nM) comparable to that of AngII ($K_d$ = 0.9 nM) whereas cyclic analogs [Sar$^1$Cys$^3$Hcy$^5$]AngII ($K_d$ = 324 nM) and [Sar$^1$Cys$^{3,5}$]-AngII ($K_d$ = 674 nM) showed low binding affinities (Table 1).

## 3.2 Evaluation of potency and efficacy of cyclic analogs on AT$_1$R signalling pathways

The potency and efficacy of the cyclic AngII analogs were then evaluated in 9 different signaling pathways. We first measured activation of the Gq/11 pathway by measuring IP$_1$ production (Fig. 1A). For this assay, the AngII dose-response curve revealed a maximal production of 3352 nM IP$_1$ (efficacy normalized to 100%) with a half-maximal response (EC$_{50}$) obtained at a concentration of 4 nM. The [Sar$^1$Hcy$^{3,5}$]AngII dose-response curve (Fig. 1A) showed an efficacy of 78% with a potency of 51 nM for Gq/11 activation. [Sar$^1$Cys$^3$Hcy$^5$]AngII (Fig. 1A) had an efficacy of 74% with a potency of 1.5 μM. [Sar$^1$Cys$^{3,5}$]AngII (Fig. 1A) was also a partial agonist with an efficacy of 60% and a potency of 2.9 μM. The AngII dose-response curve for G12 activation (Fig. 1B) revealed a maximal ligand-promoted BRET ratio of 0.091 (efficacy normalized to 100%) with an EC$_{50}$ of 6.6 nM. The [Sar$^1$Hcy$^{3,5}$]AngII dose-response curve (Fig. 1B) showed a full efficacy of 101% with a potency of 8.0 nM. [Sar$^1$Cys$^3$Hcy$^5$]AngII and [Sar$^1$Cys$^{3,5}$]AngII (Fig. 1B) were partial agonists with low efficacies of 37% and 42% and low potencies of 125 nM and 949 nM respectively. We next evaluated the activation of signalling pathways from the Gi family, namely Gi2, Gi3 and Gz. For Gi2, the AngII dose-response curves showed a maximal ligand-promoted BRET ratio of 0.044 with an $EC_{50}$ of 2.2 nM (Fig. 1C). The [Sar[1]Hcy[3,5]]AngII dose-response curve (Fig. 1C) showed a full efficacy of 100% with a potency of 3.7 nM. [Sar[1]Cys[3]Hcy[5]]AngII and [Sar[1]Cys[3,5]]AngII (Fig. 1C) were partial agonists with efficacies of 64% and 53% and potencies of 92 nM and 58 nM respectively. For Gi3, the AngII dose-response curves showed a maximal ligand-promoted BRET ratio of 0.099 with an $EC_{50}$ of 1.5 nM (Fig. 1D). The [Sar[1]Hcy[3,5]]AngII dose-response curve (Fig. 1D) showed a full efficacy of 100% with a potency of 2.5 nM. [Sar[1]Cys[3]Hcy[5]]AngII and [Sar[1]Cys[3,5]]AngII (Fig. 1D) showed efficacies of 92% and 73% and potencies of 62 nM and 67 nM respectively. For Gz, the AngII dose-response curves showed a maximal ligand-promoted BRET ratio of 0.061 with an $EC_{50}$ of 1.5 nM (Fig. 1E). The [Sar[1]Hcy[3,5]]AngII dose-response curve (Fig. 1E) showed a full efficacy of 97% with a potency of 7.4 nM. [Sar[1]Cys[3]Hcy[5]]AngII and [Sar[1]Cys[3,5]]AngII (Fig. 1E) showed efficacies of 70% and 64% and potencies of 29 nM and 18 nM respectively. The AngII dose-response curve for βarr1 recruitment showed a maximal ligand-promoted BRET ratio of 0.077 (efficacy normalized to 100%) with an $EC_{50}$ of 4.1 nM (Fig. 1C). The AngII dose-response curve for βarr2 recruitment showed a maximal ligand-promoted BRET ratio of 0.095 (efficacy normalized to 100%) with an $EC_{50}$ of 3.3 nM (Fig. 1C). The [Sar[1]Hcy[3,5]]AngII dose-response curves (Fig. 1C and Fig. 1D) for βarr1 and βarr2 recruitment showed full efficacies of 97% and 100% with potencies of 4.9 nM and 4.7 nM respectively. [Sar[1]Cys[3]Hcy[5]]AngII (Fig. 1C and Fig. 1D) was a partial agonist for βarr1 and βarr2 recruitment with efficacies of 46% and 70% and low potencies of 183 nM and 222 nM respectively. [Sar[1]Cys[3,5]]AngII (Fig. 1C and Fig. 1D) was also a partial agonist for βarr1 and βarr2 recruitment with efficacies of 53% and 61% and low potencies of 891 nM and 980 nM respectively. The AngII dose-response curve for PKC-dependent ERK activation (Fig. 1E), as determined in the presence of the EGFR inhibitor PD168393 revealed a maximal response of 52 703 luminescence arbitrary units (efficacy normalized to 100%) with an $EC_{50}$ of 4.8 nM. The AngII dose-response curve for EGFR-dependent ERK activation (Fig. 1F), in the presence of PKC inhibitor Go6983 revealed a maximal response of 66 302 luminescence arbitrary units (efficacy normalized to 100%) with an $EC_{50}$ of 3.2 nM. The [Sar[1]Hcy[3,5]]AngII dose-response curve (Fig. 1E and Fig. 1F) for PKC-dependent and EGFR-dependant ERK activation showed a full efficacy of 107% and 99% with a potency of 13 nM and 10 nM respectively. [Sar[1]Cys[3]Hcy[5]]AngII and [Sar[1]Cys[3,5]]AngII (Fig. 1E) were partial agonists of the PKC-dependent ERK activation with efficacies of 72% and 55% with low potencies of 234 nM and 280 nM respectively. [Sar[1]Cys[3]Hcy[5]]AngII and [Sar[1]Cys[3,5]]AngII (Fig. 1F) were also partial agonists of the EGFR-dependent ERK activation with efficacies of 85% and 62% with low potencies of 299 nM and 187 nM respectively. All these results are summarized in Table 2.

### 3.3 Quantification of ligand bias

#### 3.3.1 'Within-pathway' comparison of ligand efficacies

Variations in the potencies and efficacies of AngII cyclic analogs towards the different signaling pathways were observed (Table 2). This suggests the presence of signaling bias. In order to clearly establish whether an analog was biased towards one pathway over the others, the bias factors were determined for each analog and for all the signaling pathways studied. Transduction ratios [$\log(\tau/K_A)$] for the cyclic analogs were first derived using the operational model. The $\log(\tau/K_A)$ of the reference compound AngII was then subtracted from the $\log(\tau/K_A)$ value of each analog for a given pathway, yielding $\Delta\log(\tau/K_A)$ as a '*within-pathway*' comparison for each signaling pathway. The $\Delta\log(\tau/K_A)$ value is indicative of how well a given signaling pathway can be activated by a ligand, where a value of 0 indicates that a given ligand activates a pathway to the same degree as the reference compound, a positive value indicating that the ligand more strongly activates the signaling pathway than the reference compound and an increasingly negative value indicating that the ligand poorly activates the signaling pathway, if at all. Using these values, every signaling pathway was represented on a radar plot, adapted from the 'web of efficacy [35,36] (Fig.2). The $\log(\tau/K_A)$ mean value of AngII for each pathway evaluated was 8.39. [Sar[1]Hcy[3,5]]AngII had $\log(\tau/K_A)$ values ranging from 7.92 to 8.33, with $\Delta\log(\tau/K_A)$ values ranging from -0.62 to 0.01, for each pathway evaluated with the exception of the $G_q$ pathway (Table 3 and Fig 2A). For the $G_q$ pathway, the calculated $\log(\tau/K_A)$ was 6.79, with a $\Delta\log(\tau/K_A)$ of -1.70, indicating a decreased capacity of this analog to activate the $G_q$ pathway (Table 3 and Fig 2A). As for [Sar[1]Cys[3]Hcy[5]]AngII (Fig. 2B) and [Sar[1]Cys[3,5]]AngII (Fig. 2C), $\log(\tau/K_A)$ values ranged from 4.76 to 6.27 in all of the signaling pathways, with $\Delta\log(\tau/K_A)$ values ranging from -3.30 to -2.10 (Table3), indicating that both analogs were poor activators of every signaling pathway evaluated.

*3.3.2 Quantification of 'between-pathway' bias*

Ultimately, a '*between-pathway*' comparison was achieved for a given ligand in the form of $\Delta\Delta\log(\tau/K_A)$, which led to the actual bias factor ($10^{(\Delta\Delta\log(\tau/K_A))}$) (Tables 4 and 5). Table 4 shows that the bias of [Sar[1]Hcy[3,5]]AngII toward βarr1 and βarr2 over Gq/11 was 36-fold and 44-fold respectively. The bias of [Sar[1]Hcy[3,5]]AngII toward PKC-ERK and EGFR-ERK was 20-fold and 12-fold respectively whereas the bias toward G12 was 51-fold. For [Sar[1]Cys[3]Hcy[5]]AngII and [Sar[1]Cys[3,5]]AngII the bias factors were less pronounced.

### 3.3.3 Effect on rank order of potency

We have performed a rank order of potency analysis for the affinity as well as for each pathway presented in this study (Table 6). Our results show that the endogenous reference ligand AngII was always the most powerful agonist for all signaling pathways examined. [Sar[1]Hcy[3,5]]AngII was the most powerful macrocyclic AngII analog of the described series and shows a potency equal to the AngII in all of the signaling pathways except for Gq/11, and ERK were this ligand is less potent than the reference compound. [Sar[1]Cys[3]Hcy[5]]AngII and [Sar[1]Cys[3,5]]AngII are usually ranked in this order for both affinity and signaling activity except for three signaling pathways, Gz, βarr1, and G12. For Gz and βarr1 pathways, both ligands are shown to be equally potent however, there is an inversion in the rank order of potency for the G12, [Sar[1]Cys[3,5]]AngII being more potent than [Sar[1]Cys[3]Hcy[5]]AngII to trigger a G12 response.

These rank order of potency results confirm that there is a change in agonistic properties of the macrocyclic AngII analogs leading to ligand bias and that the difference observed is not related to a loss of affinity of the ligands at AT1R.

### 3.4 Molecular dynamics simulations of liganded-AT$_1$R complexes

#### 3.4.1 Modeling of ligand structures

In order to better understand how the conformational restriction conferred by the intramolecular disulfide bond of the AngII cyclic analogs affects the ligand binding conformation and the conformational landscape of the receptor, we performed molecular dynamic (MD) simulations. Initially, we modeled the structure of all three cyclic ligands in water (Fig. 3A) in a 10 ns MD simulation. These simulations were then used to evaluate the distance between the C$_\beta$ atoms of the residues in positions 3 and 5 (Fig. 3B). The MD simulation of [Sar$^1$Hcy$^{3,5}$]AngII showed an average distance between the C$_\beta$ atoms of Hcy$^3$ and Hcy$^5$ of $0.54 \pm 0.05$ nm. The corresponding distances were $0.48 \pm 0.03$ nm in the MD simulation of [Sar$^1$Cys$^3$HCy$^5$]AngII and $0.42 \pm 0.03$ nm in the MD simulation of [Sar$^1$Cys$^{3,5}$]AngII (Fig. 3B), which were below the distances evaluated for AngII. This suggests that the cyclic analogs have a more constrained structure in water when compared to AngII.

#### 3.4.2 MD simulations of the AngII-AT$_1$R complexes

We next generated a complex between AngII and AT$_1$R by docking our previously established model of AngII bound to a homology model of the AT$_1$R [17,37] onto the three-dimensional crystal structure of the AT$_1$R [2]. One microsecond of MD simulation time was initially performed on the AngII-AT$_1$R complex in the form of ten trajectories each 100 ns in length. The trajectories were analyzed to assess whether AngII bound to the AT$_1$R natively adopts conformations compatible with cyclization between residues 3 and 5. To do so, we monitored the distribution of the distance between the $C_\beta$ atoms of Val[3] and Ile[5] (Fig. 4).  Our analysis revealed that the distance between the $C_\beta$ atoms ranged from 0.55 nm to 0.92 nm (Fig. 4A). Furthermore, we observed that AngII adopts a conformation where this distance is 0.62 nm or less, permissible for the cyclization of [Sar[1]Hcy[3,5]]AngII, during 4.9 % of the simulation. However, for both [Sar[1]Cys[3]HCy[5]]AngII and [Sar[1]Cys[3,5]]AngII, the required distance for cyclization must be shorter. This result indicates that AngII can spontaneously adopt conformations that bring the side-chains of Val[3] and Ile[5] in close proximity and that cyclization of AngII through these positions will not lead to non-native conformations. To verify that the relatively short simulation time of 1 μs did not itself introduce bias, the AngII-AT$_1$R complex simulation was thus prolonged to 5 μs (10 trajectories of 500 ns) to further sample the conformational space. Prolonging the simulations did not reveal any new populations, hence we feel comfortable that the short simulation time was sufficient to obtain a meaningful trajectory. Our results suggest that the disulfide bond in [Sar[1]Hcy[3,5]]AngII biases the AT$_1$R conformational space towards structures that are less capable of activating the Gq pathway. We therefore performed the same simulation using [Sar[1]Ile[8]]AngII, which does not activate Gq upon binding the AT$_1$R [19].The probability distribution of the [Sar[1]Ile[8]]AngII-AT$_1$R simulation shows a small population optimum at $d = 0.58$ nm, with a second configuration with an optimum at $d = 0.82$ nm, as also seen in the AngII-AT$_1$R simulation. During the [Sar[1]Ile[8]]AngII-AT$_1$R simulation, we observed that the distance between the $C_\beta$ of Val[3] and Ile[5] was less than 0.62 nm for 17.0 % of the simulation, compared to 4.9 % in the AngII-AT$_1$R simulation.

### 3.4.3 MD simulations of the [Sar[1]Cys[3]Hcy[5]]AngII-AT$_1$R complexes

To further document the fact that the conformation of the cyclic [Sar[1]Hcy[3,5]]AngII is detrimental to the stabilization of $AT_1R$ conformations capable of activating the Gq pathway, we performed 3 molecular dynamics simulations of the [Sar[1]Hcy[3,5]]AngII-$AT_1R$ complex with the cyclic ligand (#C1, #C2 and #C3) together with three AngII-$AT_1R$ control simulations in which the ligands #L1, #L2 and #L3 have the same corresponding starting positions and backbone conformations as #C1, #C2 and #C3 but where Hcy[3] and Hcy[5] where replaced by the native Val[3] and Ile[5]. The distance between the $C_\beta$ atoms of Val[3] and Ile[5] (or Hcy[3] and Hcy[5]) were measured in each simulation (Fig. 4B). Predictably, the [Sar[1]Hcy[3,5]]AngII-$AT_1R$ simulations display a narrow distance distribution. Simulations #C1 and #C2 feature two population optima at d = 0.54 nm and 0.60 nm and simulation #C3 has a single optimum at 0.58 nm. The corresponding simulations of the parent AngII-$AT_1R$ complexes (# L1, #L2 and #L3) indicate that AngII has the capacity to adopt a variety of potentially stable conformations and maintain native conformations with a distance of $\leq 0.62$ nm between the $C_\beta$ atoms of Val[3] and Ile[5], which was not hindered by initial placement of the ligand in the simulations.

We examined more precisely the conformations adopted by AngII with a focus on residues 3 to 5 and noted three recurrent configurations (Fig. 5). We observed a β-strand conformation for Val[3], Tyr[4] and Ile[5] of AngII with the distance of about 0.70 nm between the $C_\beta$ of Val[3] and Ile[5] as expected in a typical anti-parallel β-sheet. We also observed a γ-turn conformation of the backbone which allows for the positioning of the $C_\beta$ atoms of Val[3] and Ile[5] within the distance required for the formation of the disulfide bond between two homoCys residues. Lastly, an extended backbone conformation with a larger distance (0.8 nm) was also observed.

*3.5 Environments explored by the C-terminal Phe[8] of AngII and [Sar[1]Hcy[3,5]]AngII within the AT$_1$R*

To further document the structural and dynamical differences between AngII and [Sar[1]Hcy[3,5]]AngII in complex with the AT$_1$R we monitored the location and the environment of the C-terminal Phe[8] residue in each simulation. Phe[8] is known to be an crucial determinant for AT$_1$R Gq signaling and its substitution with hydrophobic residues leads to βarr biased ligands, as is the case with [Sar[1]Ile[8]]AngII [19,38,39]. We had previously described [37] that the insertion of Phe[8] within the hydrophobic core, which we ascribed to the activation of Gq, occurs when the distance between Phe[8] of AngII and residue F77[2.53] of the AT$_1$R is approximately 0.86 nm or less (Fig. 6). The distance between the side chain of Phe[8] and the side chain of F77[2.53] of the AT$_1$R was thus monitored for all simulations (Fig. 7). The simulation of the AngII-AT$_1$R complex showed a wide distribution of distances with peaks at $d$ = 0.72 nm (Phe[8] inside the hydrophobic core) and $d$ = 1.10 nm (Phe[8] outside the hydrophobic core) (Fig. 7A). The distance was 0.86 nm or less for 37.6 % of the AngII-AT$_1$R simulation time. The simulation of the AngII-AT$_1$R complex was also prolonged to 5 μs (10 trajectories of 500 ns) from the initial 1 μs (Fig. 7A). The populations with Phe[8] positioned in the hydrophobic core (at $d$ = 0.70 nm) or outside the hydrophobic core (at $d$ = 1.10 nm) show similar probabilities but are more well defined than in the initial simulation of the AngII-AT$_1$R complex. We also performed a simulation using [Sar[1]Ile[8]]AngII, that is incapable of activating Gq. This simulation shows that position 8 only rarely inserts into the hydrophobic core (Fig. 7A).

Simulations in #C1 and #L1 showed (Fig. 7B) that the side-chain of Phe[8] of the linear peptide is more likely to be in the hydrophobic core than that of the cyclic peptide. Simulations #L2, #C2, #L3 and #C3 all favored a distance close to their respective starting configurations (Fig. 7B). Interestingly, the probability distribution of #L2 indicates Phe[8] inserted in the hydrophobic core on occasion ($d \leq 0.86$ nm for 7.5 % of the simulation) while it did not in simulation #C2 ($d \leq 0.86$ nm for 0.1 % of the simulation). The probability of Phe[8] being inserted in the hydrophobic core was slightly higher with the cyclic peptide in simulation #C3 than with the linear peptide in #L3 ($d \leq 0.86$ nm for 97.1 % in #C3 and 95.1% in #L3). Combining all three simulations of each ligand, our analysis shows that AngII has an increased probability of sampling a transition state (centered at $d \approx 0.90$ nm on the probability distribution) than does [Sar[1]Hcy[3,5]]AngII (Fig 7C). The proportion of total population with $d \leq 0.86$ nm was also larger when comparing the three linear (#L1, #L2 and #L3) simulations (46.7 %) to the three cyclic (#C1, #C2 and #C3) simulations (42.5 %). Taken together, the results indicate that [Sar[1]Hcy[3,5]]AngII maintains a stable conformation with Phe[8] inserted in the hydrophobic core, but both this state and the transition state are sampled less often than with AngII.

*3.6 2D probability distribution functions reveals a link between the conformation of the backbone of residues Val[3] to Ile[5] and the position of side-chain of Phe[8] in the $AT_1R$*

In order to evaluate the possible correlation between the structure of the ligand and the position of Phe[8] relative to the hydrophobic core, we generated two-dimensional (2D) probability distribution functions (PDFs). We obtained PDFs where the distance between Phe[8] and F77[2.53] of the AT$_1$R is plotted on the x axis and the distance between the C$_\beta$ of the residues in position 3 and 5 of the ligand is plotted on the y axis (Fig. 8). All initial ligand positions are denoted by a star. To facilitate interpretation, 2D PDFs were divided into quadrants. The simulation of the AngII-AT$_1$R complex indicates that the ligand can adopt γ-turn ($d < 0.70$ on the *y* axis, *bottom quadrants*), β-strand ($d = 0.70$ on the *y* axis) or extended conformations ($d > 0.70$ on the *y* axis, *top quadrants*) with Phe[8] either inserted within the hydrophobic core ($d < 0.86$ on the *x* axis, *left quadrants*) or excluded from the hydrophobic core ($d > 0.86$ on the *x* axis, *right quadrants*) (Fig. 8). For the AngII-AT$_1$R complex, we observed that there was a lower probability (dark blue) of AngII to sample a distance of $d = 0.86$ nm on the *x* axis when in a γ-turn conformation (*bottom quadrants*) compared to the higher probability (yellow or green) when AngII is in an extended conformation (*top quadrants*). This indicates that the transition of Phe[8] inside or outside of the hydrophobic core is facilitated when the ligand is in an extended conformation compared to the γ-turn conformation. The conformational landscape of the prolonged simulation of the AngII-AT$_1$R complex (5 µs) shows one additional small population with $d > 1.5$ nm on the *x* axis describing a state where Phe[8] of AngII is outside the hydrophobic core. The prolonged 2D probability distribution also supports the observations derived from the 1 µs simulations. When we examined the conformational landscape from the simulation of the [Sar[1]Ile[8]]AngII-AT$_1$R complex (Fig. 8) we found that the extended configuration of the ligand with Ile[8] positioned outside of the hydrophobic core is favored (*top right quadrant*). Well-defined populations of the ligand in γ-turn conformation are observed with Ile[8] positioned both in (*bottom left quadrant*) and out (*bottom right quadrant*) of the hydrophobic core. Furthermore, the conformational landscape shows the γ-turn conformation of the ligand (*bottom left quadrant*) is preferred over the extended conformation (*top left quadrant*) when the distance between Phe[8] and F77[2.53] is below the 0.86 nm threshold.

Simulation #C1 of the cyclic ligand shows that the γ-turn conformation of the ligand is favored. The 2D PDF also shows that the most stable conformation of [Sar[1]Hcy[3,5]]AngII, is observed when Phe[8] is inserted in the hydrophobic core (*bottom left quadrant*). We also observed two populations outside of the hydrophobic core (*bottom right quadrant*) that correspond to the two ligand conformers shown on the one-dimensional probability distributions (Fig. 3B). These two conformers have equal population, and correspond to ligands that have intermediate conformations between the γ-turn and the β-strand. The corresponding simulation #L1 using the linear peptide shows a preferred conformation of the ligand close to its initial conformation (*bottom left quadrant*), which is compatible with the conformation of the cyclic ligand (Fig. 8). A minor population where the ligand maintained Phe[8] close to F77[2.53] while adopting an extended conformation can also be observed (*top left quadrant*). The landscape shows the γ-turn conformation of AngII is preferred when Phe[8] is outside of the hydrophobic core (*bottom right quadrant*). These simulations illustrate that a conformationally constrained AngII ligand can still access the AT[1]R hydrophobic core leading to Gq activation. The conformational landscape of simulation #C2 displays a single population with Phe[8] positioned outside of the hydrophobic core (*bottom right quadrant*) while that of the corresponding control simulation #L2 shows a similar population but shifted upwards on the *y*-axis. Simulation #C3 displays a single population with Phe[8] inside the hydrophobic core (*bottom left quadrant*) (Fig. 8). The corresponding control simulation #L3 shows a population distribution that is wider and shifted upwards on the *y*-axis compared to #C3, centered close to the β-strand conformation but also sampling both the γ-turn and extended conformations (Fig. 8). A minor population is observed with Phe[8] being out of the hydrophobic core and the ligand being in extended configuration (*top right quadrant*). For both sets of simulations #C2/#L2 and #C3/#L3, the favored populations of the ligand were very close to their initial conformation, and these ligands were only able to transition into (#C2/#L2) or out of (#C3/#L3) the AT$_1$R hydrophobic core with less probability. The results of our simulations support the notion that the γ-turn conformation forced by the homoCys-homoCys disulfide bond present in [Sar[1]Hcy[3,5]]AngII lowers the probability of insertion of Phe[8] into the hydrophobic core compared to unconstrained AngII.

## 4. DISCUSSION

Ligand bias or functional selectivity is a validated concept that is becoming increasingly relevant in the field of pharmacology. It allows for the fine tuning of a cellular response by enabling a receptor to adopt a conformation that leads to the activation of specific signaling pathways. Development of ligands able to elicit bias could introduce selective therapeutic approaches for desired targets by activating pathways responsible for beneficial effects without activating pathways that could possibly yield undesired ones [13–15]. In this vein, the purpose of this study was to investigate whether constraining the conformation of AngII by using analogs that were cyclized at positions 3 and 5 would lead to ligand bias and, using molecular dynamics simulations, to understand how such analogs interact with the $AT_1R$ to allow such a bias.

We determined that the cyclic ligand [$Sar^1Hcy^{3,5}$]AngII has mild negative bias against the Gq pathway, as shown in Fig. 2. Indeed, among the pathways tested, the potency and efficacy of this ligand were reduced exclusively on this pathway. As for cyclic ligands [$Sar^1Cys^3Hcy^5$]AngII and [$Sar^1Cys^{3,5}$]AngII, where the ring size is reduced respectively by 1 and 2 atoms (Fig. 3A), the efficacy and the potency were significantly reduced in all pathways evaluated and these ligands showed no significant bias in any pathway tested. Previous work has demonstrated that [$Sar^1Ile^8$]AngII acts as a biased agonist for G12, βarrs and ERK pathways while not activating Gq [19]. Furthermore, both [$Sar^1Hcy^{3,5}$]AngII and [$Sar^1Ile^8$]AngII exhibited one of the advantages of functional selectivity of the $AT_1R$ by only activating the signaling pathway associated with cardiac benefits (βarr) without fully activating Gq, responsible for less desirable hypertensive effects [15].

The molecular dynamic simulations (Fig. 3B, #L1) show that AngII, in our model of the AngII-AT$_1$R complex, has the capacity to adopt a stable γ-turn conformation between residues Val$^3$ and Ile$^5$ which is very similar to the conformation of the cyclic ligand [Sar$^1$Hcy$^{3,5}$]AngII. However, in the other two control simulations (Fig 9B, #L2 and #L3), AngII only transiently adopted this conformation. Thus, while AngII can adopt and maintain the γ-turn conformation for residues Val$^3$ through Ile$^5$ while bound to AT$_1$R, more extended conformations seem favored. As for the other two cyclic analogs, [Sar$^1$Cys$^3$Hcy$^5$]AngII and [Sar$^1$Cys$^{3,5}$]AngII, the distance between the C$_\beta$ atoms of the residues in positions 3 and 5 was too long for those analogs to adopt conformations that were compatible with AngII. This could explain why the binding affinity of cyclic ligands decreases as the number of atoms forming the cycle diminishes. It would be interesting to further investigate whether increasing the size of the cycle could lead to compounds with better affinities or functional selectivity profiles.

Monitoring the position of Phe$^8$ of AngII within the binding pocket revealed that [Sar$^1$Hcy$^{3,5}$]AngII is at a disadvantage when compared to AngII to insert the aromatic sidechain of Phe$^8$ within the hydrophobic core. This could explain the reduced potency and efficacy of [Sar$^1$Hcy$^{3,5}$]AngII on the Gq pathway. The importance of Phe$^8$ of AngII for Gq-mediated signaling is well documented and we have previously linked its insertion within the hydrophobic core to the activation of the Gq pathway [19,37,40]. The data obtained in the current study suggest that the energy barrier that needs to be overcome for the insertion of Phe$^8$ in the hydrophobic core is increased in [Sar$^1$Hcy$^{3,5}$]AngII compared to AngII, possibly due to the conformational restriction imposed by the disulfide bond. This barrier also appears increased for AngII and [Sar$^1$Ile$^8$]AngII when these linear ligands adopt a γ-turn conformation (similar to the cyclic ligand) compared to the extended conformation. Thus, when compared to the β-strand and extended conformations, the γ-turn conformation of AngII, to which the cyclic analog is confined, is less effective at stabilizing the Gq-active state.

It must be stressed that in our simulations, the transition of the Phe[8] residue from the exterior to the interior the hydrophobic core essentially occurred only with AngII, as the starting configurations of simulation #C1 and #C3 were generated from simulations of AngII in which it adopted a conformation compatible with the cyclic ligand. Because all simulations started with the ligand positioned within the binding pocket, the actual process of ligand binding has been bypassed, which nullifies our ability to sample potential events of conformational selection that occurs before the ligand binds the receptor. Further MD simulations will be required to better evaluate the binding and stabilization of the different states of the receptor by the ligands. Furthermore, residue Phe[8] interacts directly with the side-chain of the conserved residue W253[6.48], the "tryptophan toggle switch", which has been identified as important for G protein-mediated signaling by several GPCRs but shown to be of little importance for inositol phosphate production by the $AT_1R$ [41,42]. Its role in the activation of other pathways by the $AT_1R$ is to be determined. Likewise, another potentially conserved switch in GPCR activation, F249[6.44], is located just one helix turn below W253[6.48] but its role in $AT_1R$-mediated signaling is currently unclear. Further work is required to identify and characterize how these potential switches affect $AT_1R$ signaling and the conformational landscape explored by the receptor. The mildly biased [Sar[1]Hcy[3,5]]AngII ligand should prove to be a useful tool for this endeavor as it complements the unbiased ligand AngII and the strongly biased ligand [Sar$^1$Ile$^8$]AngII, which is completely inactive on the Gq pathway.

In conclusion, our results suggest that constraining AngII through a linkage between positions 3 and 5 by Hcy disulfide bond leads to a local γ-turn conformation that introduces a Gq pathway negative bias while retaining full agonism for all other pathways. This opens up new strategies in order to develop new biased non-peptidic analogs of the AT$_1$R that could serve in more targeted treatment and furthers our understanding of the AT$_1$R activation process.

## ACKNOWLEDGEMENT

We would like to thank Marie-Reine Lefebvre for her assistance with peptide synthesis, purification and characterization. This work was supported by a grant from the Canadian Institutes of Health Research [MOP-136770].

The authors declare no competing interests.

**FIGURE**

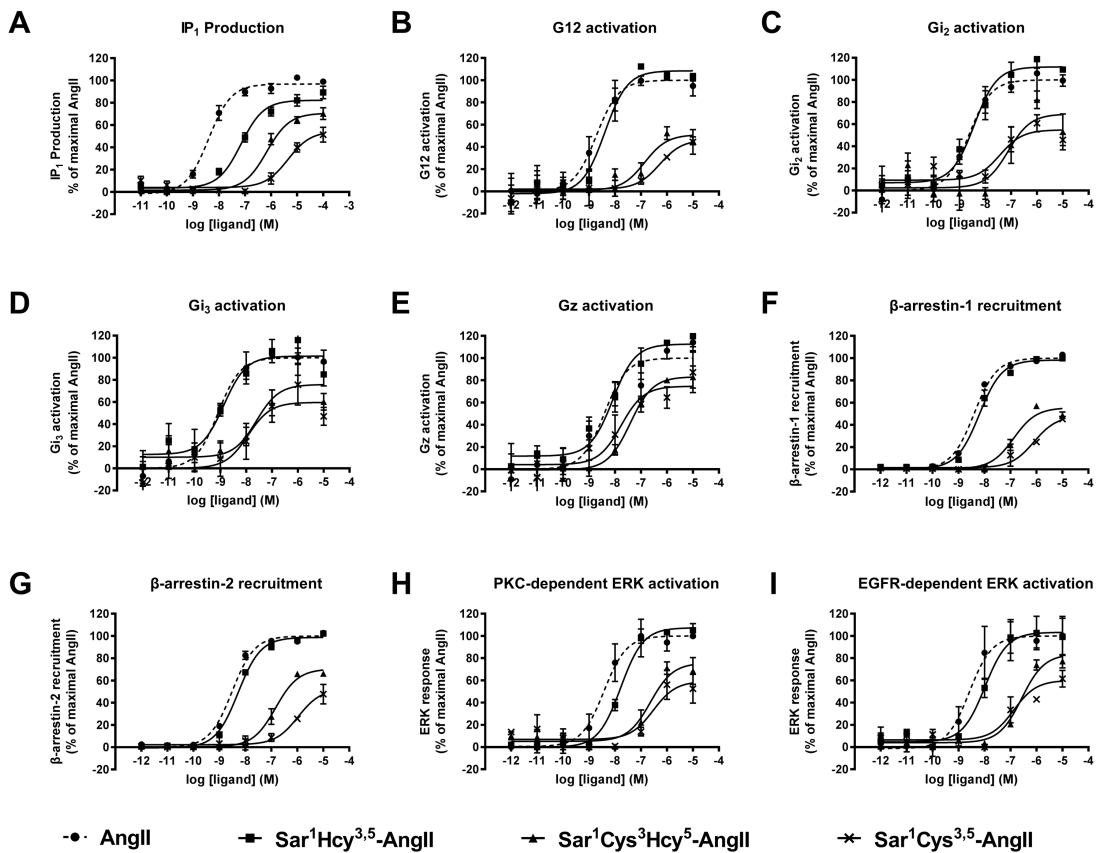

● AngII   ■ Sar[1]Hcy[3,5]-AngII   ▲ Sar[1]Cys[3]Hcy[5]-AngII   ✕ Sar[1]Cys[3,5]-AngII

**Fig. 1. Activation of $AT_1R$ signalling pathways by AngII analogs.** For all assays, cells were stimulated with increasing concentrations of AngII, [Sar[1]Hcy[3,5]]AngII, [Sar[1]Cys[3]Hcy[5]]AngII and [Sar[1]Cys[3,5]]AngII for the indicated times. (A) HEK293 cells expressing the $AT_1R$ were stimulated for 30 min at 37°C. $IP_1$ accumulation was measured with the IP-One assay, as described in the methods. (B) HEK293 cells co-transfected with $AT_1R$, $G\alpha12$-RLucII, $G\gamma1$-GFP10 and $G\beta1$ were stimulated for 8 min at 37°C. G12 activity was measured as described in the methods. (C, D) HEK293 cells co-transfected with $AT_1R$, $G\alpha i2$-RLucII (C) or $G\alpha i3$-RLucII (D), $G\gamma2$-GFP10 and $G\beta1$ were stimulated for 1 min at 37°C. (E) HEK293 cells co-transfected with $AT_1R$, $G\alpha z$-RLucII, $G\gamma1$-GFP10 and $G\beta1$ were stimulated for 5 min at 37°C. (F, G) HEK293 cells co-transfected with fusion protein

RLucII-βarr1 (F) or RLucII-βarr2 (G) together with $AT_1R$-GFP10 were stimulated for 8 min at 37°C. βarr recruitment was measured as described in the methods. (H) HEK293 cells expressing the $AT_1R$ were pretreated with 250 nM PD168393 for 30 min and then were stimulated for 2 min at 37°C. ERK activity was measured as described in the methods. (I) HEK293 cells expressing the $AT_1R$ were pretreated with 1 uM Go6983 for 30 min and then were stimulated for 5 min at 37°C. ERK activity was measured as described in the methods. Data are expressed as a percentage of AngII maximal response. Data are the mean ± SD of 3-6 independent experiments performed in triplicate.

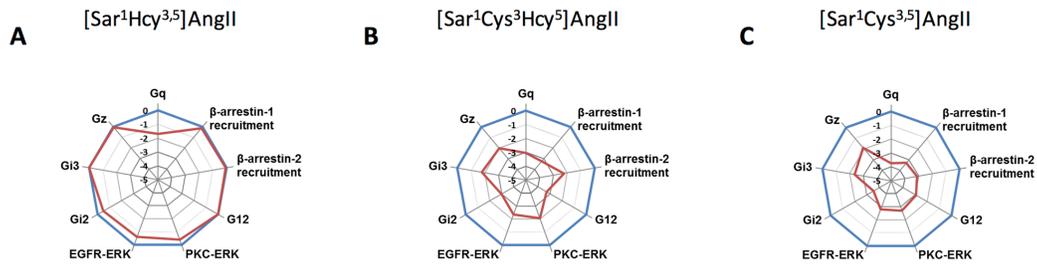

**Fig. 2. Effects of AngII analogs on AT$_1$R signaling pathways.** Radar graph representations summarizing the calculated $\Delta\log(\tau/K_A)$ values of the different ligand-activated pathways [Sar$^1$Hcy$^{3,5}$]AngII (A), [Sar$^1$Cys$^3$Hcy$^5$]AngII (B) and [Sar$^1$Cys$^{3,5}$]AngII (C). The balanced reference analog AngII is represented in blue.

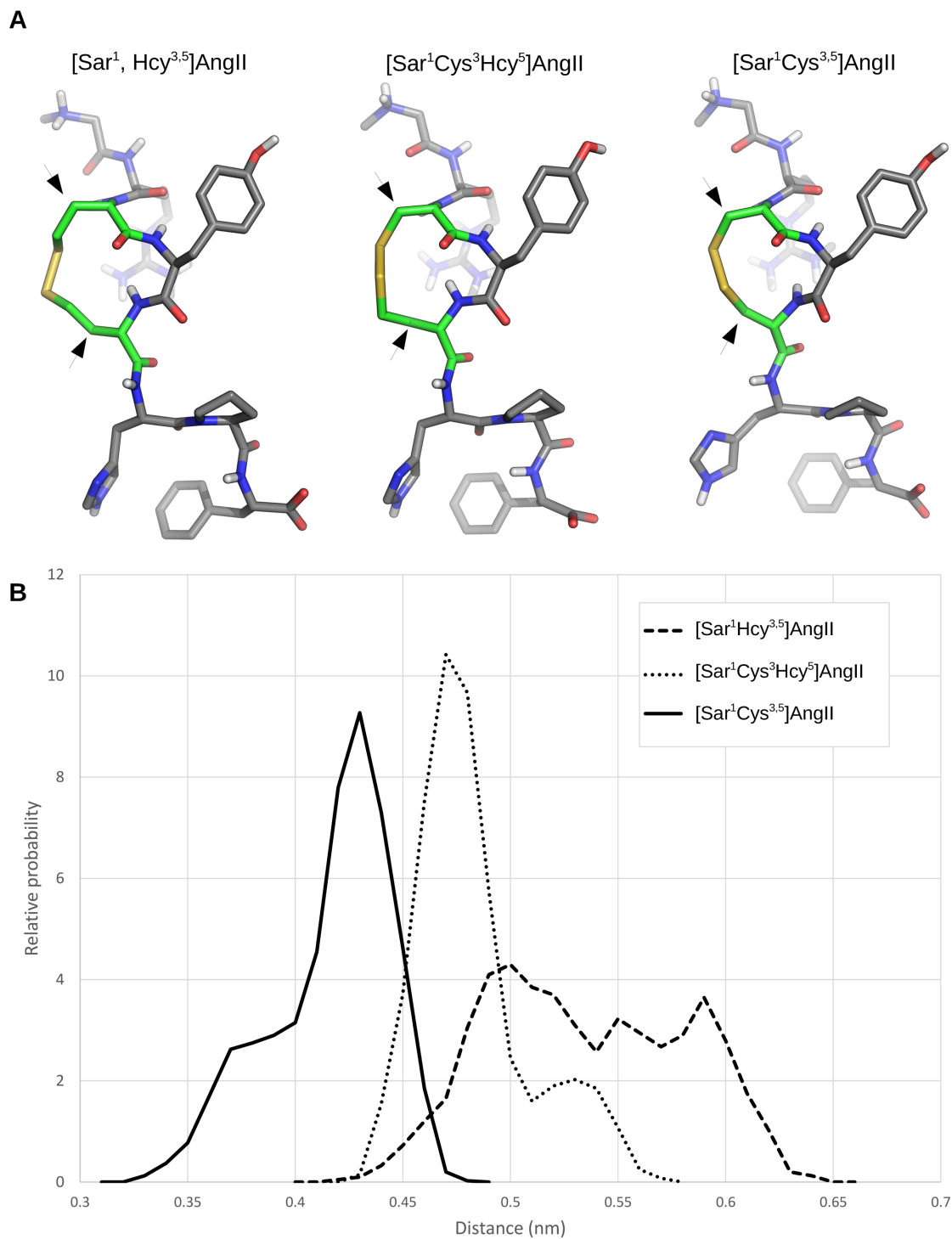

Fig. 3. (A) 3D representation of the three cyclic analogs: [Sar¹Hcy³,⁵]AngII, [Sar¹Cys³Hcy⁵]AngII and [Sar¹Cys³,⁵]AngII. Arrows indicate the Cβ atoms of residues

in positions 3 and 5. (B) Probability distribution of the distance between the C$\beta$ atoms of residues in positions 3 and 5 of the cyclic peptides in water measured from 10 ns MD simulations.

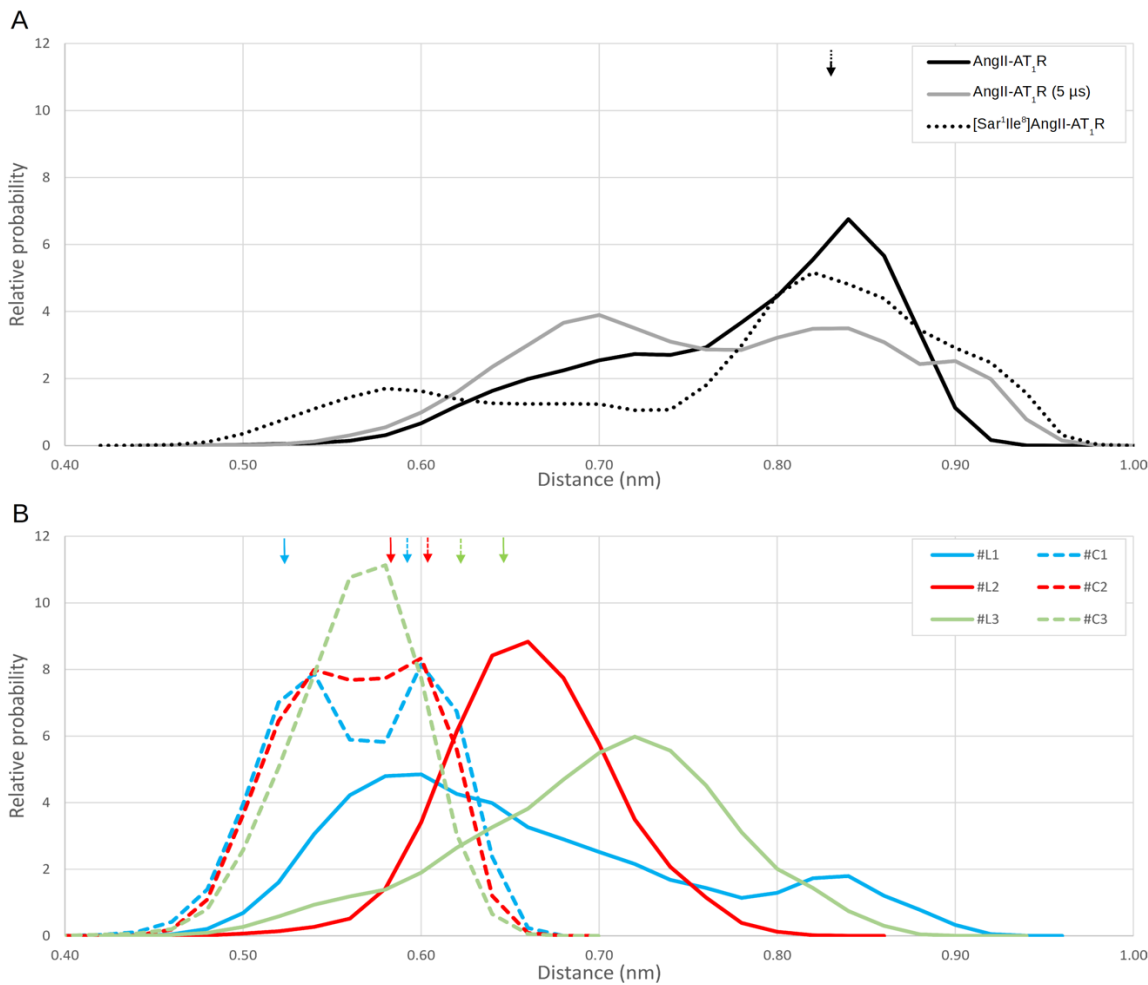

**Fig. 4. Probability distribution of the distance between the Cβ atoms of residues in positions 3 and 5 for ligands in complex with AT$_1$R measured from 1 μs MD simulations.** (A) AngII-AT$_1$R, AngII-AT$_1$R (5 μs) and [Sar$^1$Ile$^8$]AngII-AT$_1$R. (B) [Sar$^1$Hcy$^{3,5}$]AngII-AT$_1$R (#C1, #C2 and #C3) and the corresponding control simulations with AngII-AT$_1$R (#L1, #L2 and #L3). The arrows indicate the initial distance at the start of each MD simulation. The initial distance was the same for all three simulations in (A). Each simulation is composed of the trajectories of 100 ns in length.

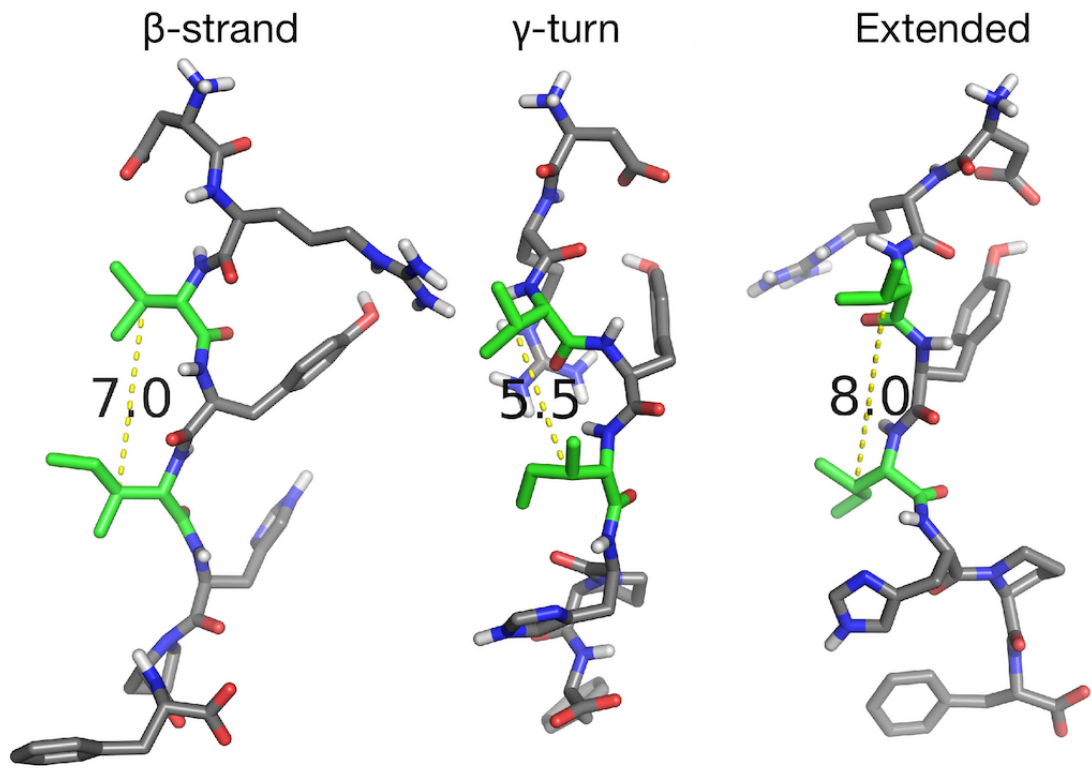

**Fig. 5. Examples of the β-strand, γ-turn and extended conformations of AngII.** The β-strand and γ-turn conformations shown are snapshots taken from MD simulations of the AngII-AT$_1$R complex. The extended conformation was manually generated for comparison purpose. Carbon atoms are colored gray, except for residues Val$^3$ and Ile$^5$ which are colored green. Oxygen atoms are red, nitrogen atoms are blue and polar hydrogens are white.

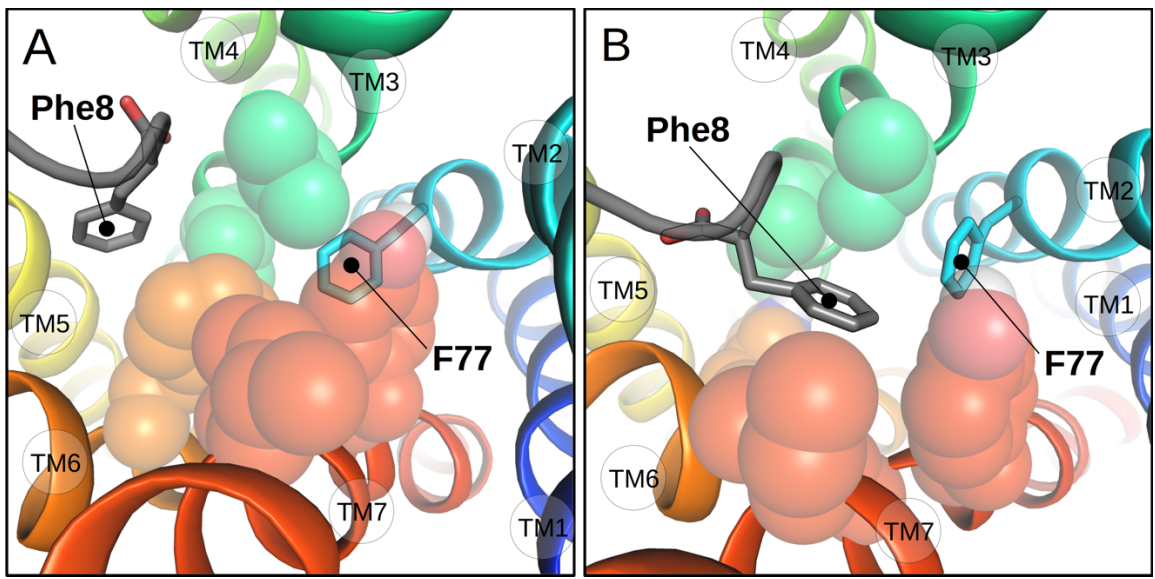

**Fig. 6. Snapshots from the MD simulation of the AngII-AT$_1$R complex showing the positioning of Phe[8] of AngII outside of the AT$_1$R hydrophobic core, further from residue F77$^{2.53}$ (A) and inside the hydrophobic core, closer to residue F77$^{2.53}$ (B).** Transmembrane domains are shown as colored ribbons and colored differently from blue (TM1) to red (TM7). The backbone of AngII is shown as a grey ribbon. The sidechains of Phe[8] and F77$^{2.53}$ are shown as sticks. Sidechains from residues forming the hydrophobic core are shown as semi-transparent spheres.

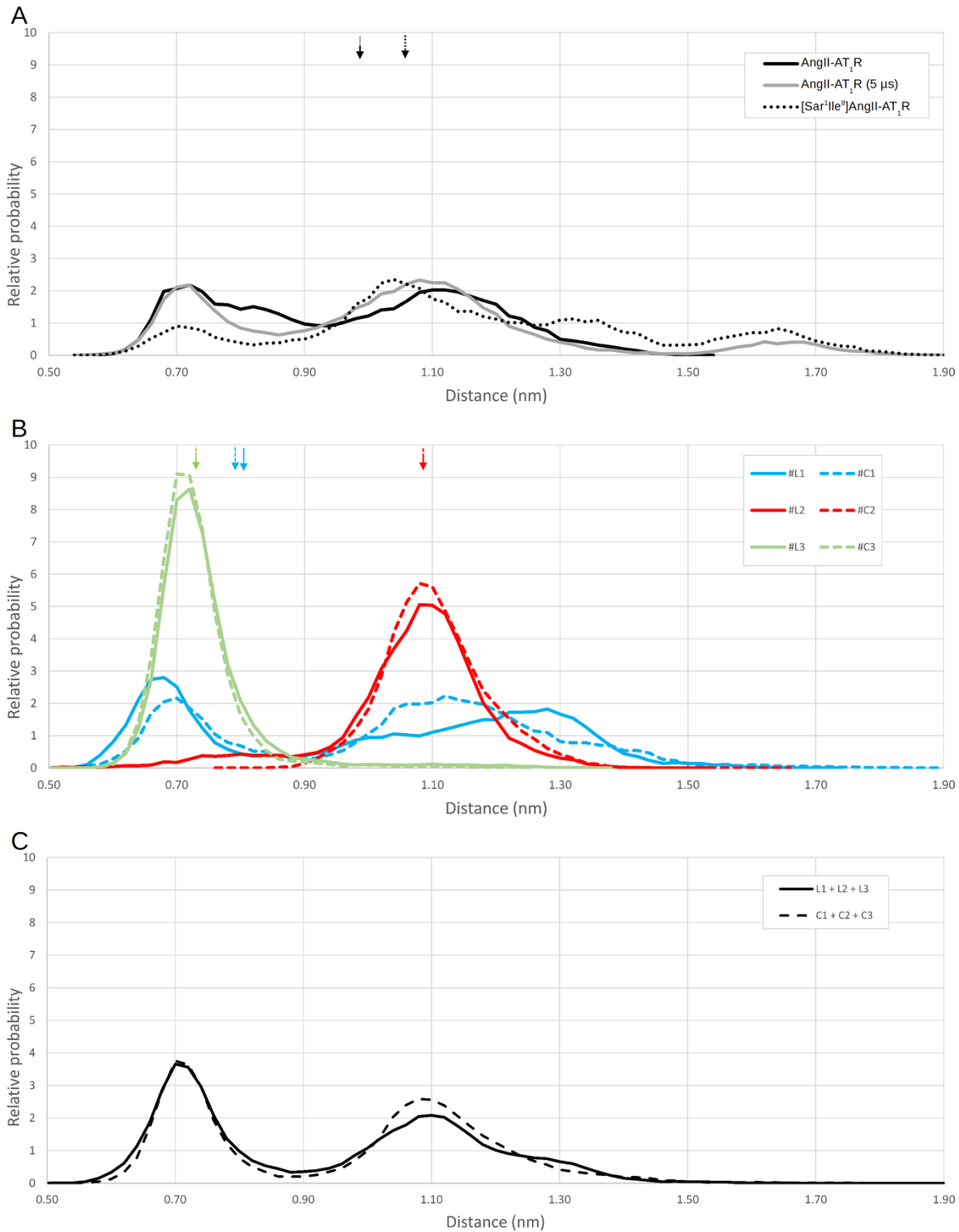

**Fig. 7. Probability distribution of the distance between the side chain of residue Phe[8] of the ligand and the side chain of residue F77[2.53] of the AT$_1$R measured from 1 μs**

**MD simulations.** (A) AngII-AT$_1$R, AngII-AT$_1$R (5 µs) and [Sar$^1$Ile$^8$]AngII-AT$_1$R. (B) [Sar$^1$Hcy$^{3,5}$]AngII-AT$_1$R (dashed lines, simulations #C1, #C2 and #C3) and the corresponding control simulations with AngII-AT$_1$R (bold lines, simulations #L1, #L2 and #L3. (C) Sum of the individual simulations shown in (B), adjusted to represent the same cumulative probability.

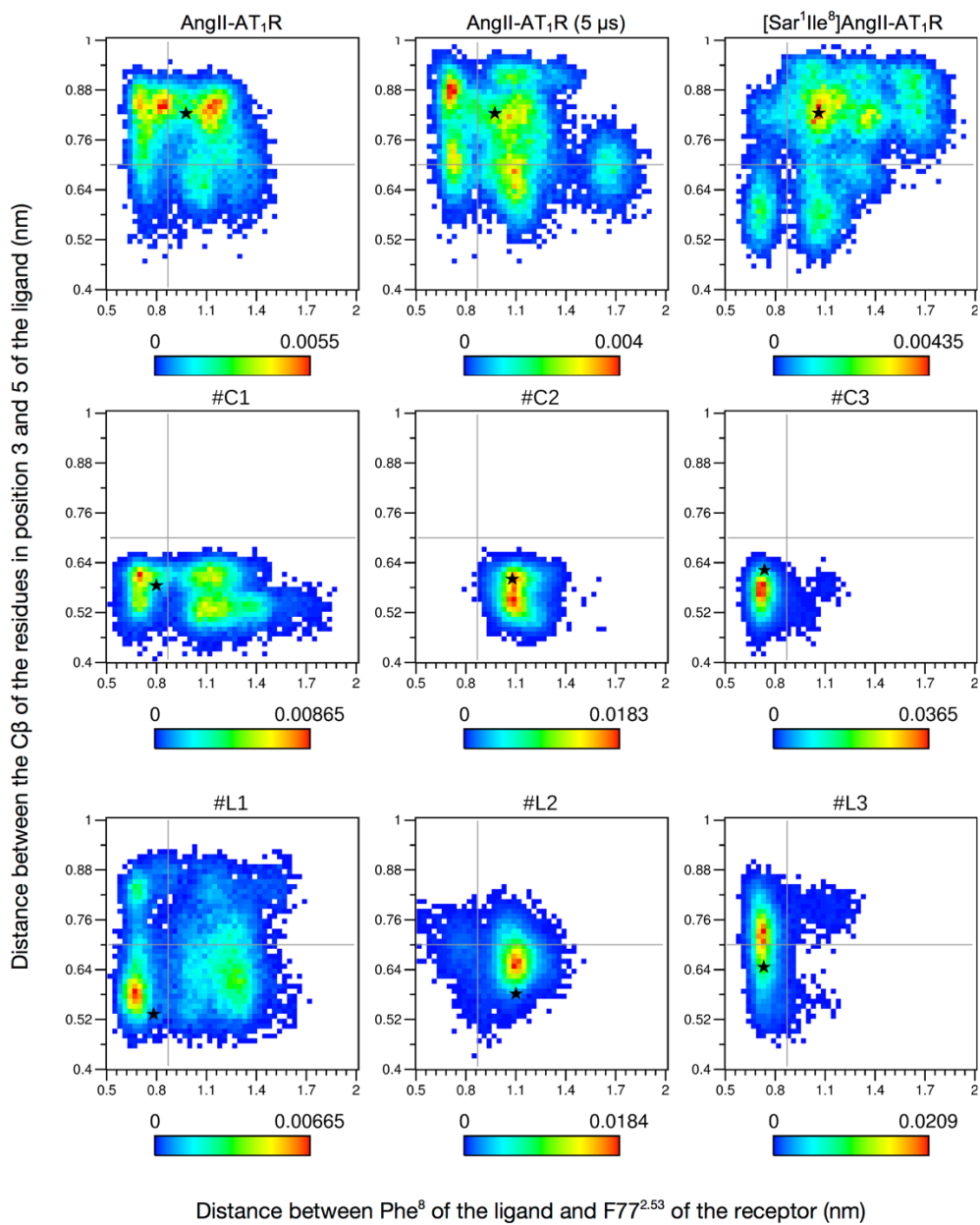

**Fig. 8. 2D probability landscapes generated by sorting frames of the MD simulations.**

(X axis) Distance between the side chain of residue Phe[8] of the ligand and the side chain of residue F77[2.53] of the AT$_1$R. The vertical grey line is positioned approximately on the transition state at $d$ = 0.88 nm as a visual aid. Populations to the left of this line are in the hydrophobic core, and those to the right of this line are outside the hydrophobic core. (Y axis) Distance between the Cβ atoms of residues in positions 3 and 5 of AngII or [Sar[1], Hcy[3,5]]AngII. The horizontal grey line is positioned at $d$ = 0.70 nm and represents the β-strand conformation of the ligand as a visual aid. Populations above this line have their ligand in the extended conformation, and those below this line have their ligand in the γ-turn conformation. The 2D probability landscapes shows AngII-AT$_1$R, AngII-AT$_1$R (5 μs) and [Sar[1]Ile[8]]AngII-AT$_1$R (first row). #C1, #C2 and #C3 represents different conformations of the [Sar[1]Hcy[3,5]]AngII-AT$_1$R complex (second row) and #L1, #L2 and #L3 represents different conformations of the control AngII-AT$_1$R complex (third row).

**Table 1**

**Binding properties of AT$_1$R ligands**

|  | K$_d$ (nM) | n |
|---|---|---|
| AngII | 0.9 ± 0.3 | 15 |
| [Sar$^1$Hcy$^{3,5}$]AngII | 2.7 ± 1.3 | 3 |
| [Sar$^1$Cys$^3$Hcy$^5$]AngII | 324 ± 128 | 3 |
| [Sar$^1$Cys$^{3,5}$]AngII | 674 ± 155 | 3 |

HEK293 cells stably expressing the AT$_1$R were assayed as described in the methods. Binding affinities (K$_d$) are expressed as the means ± SD of values obtained in *n* independent experiments performed in duplicate.

**Table 2**

**Activation of $G_q$, $G_{i2}$, $G_{i3}$, $G_z$, βarr1, βarr2, G12, PKC-ERK and EGFR-ERK by $AT_1R$ ligands**

| | AngII | | $[Sar^1Hcy^{3,5}]AngII$ | | $[Sar^1Cys^3Hcy^5]AngII$ | | $[Sar^1Cys^{3,5}]AngII$ | |
|---|---|---|---|---|---|---|---|---|
| | $EC_{50}$ (nM) | $E_{max}$ (% AngII) | $EC_{50}$ (nM) | $E_{max}$ (% AngII) | $EC_{50}$ (nM) | $E_{max}$ (% AngII) | $EC_{50}$ (nM) | $E_{max}$ (% AngII) |
| $G_{q/11}$ | 3.9 ± 1.5 | 100 | 51 ± 13** | 78 ± 6 | 1565 ± 636*** | 70 ± 5 | 2922 ± 305*** | 60 ±9 |
| $G_{i2}$ | 2.2 ± 0.7 | 100 | 3.7 ± 1.6 | 102 ± 14 | 92 ± 32** | 64 ± 6 | 58 ± 34** | 53 ± 8 |
| $G_{i3}$ | 1.5 ± 0.5 | 100 | 2.5 ± 1.5 | 109 ± 20 | 62 ± 21** | 92 ± 20 | 67 ± 19** | 73 ± 24 |
| $G_z$ | 5.6 ± 2.6 | 100 | 7.4 ± 3.8 | 97 ± 14 | 29 ± 5 | 70 ± 12 | 18 ± 7 | 64 ± 12 |
| βarr1 | 4.1 ± 0.9 | 100 | 4.9 ± 1.3 | 97 ± 2 | 183 ± 42*** | 46 ± 10 | 891 ± 155*** | 53 ± 9 |
| βarr2 | 3.3 ± 1.3 | 100 | 4.7 ± 1.0 | 100 ± 2 | 222 ± 67*** | 70 ± 2 | 980 ± 82*** | 61 ± 8 |
| $G_{12}$ | 6.6 ± 3.7 | 100 | 8.0 ± 4.6 | 101 ± 6 | 125 ± 47* | 37 ± 9 | 949 ± 475** | 42 ± 17 |
| PKC-ERK | 4.8 ± 1.6 | 100 | 13 ± 5 | 107 ± 7 | 234 ± 59*** | 72 ± 3 | 280 ± 16*** | 55 ± 4 |
| EGFR-ERK | 3.2 ± 0.7 | 100 | 10 ± 1** | 99 ± 9 | 299 ± 30*** | 85 ± 4 | 187 ± 40*** | 62 ± 9 |

$*p \leq 0.05$, $** p \leq 0.01$, and $*** p \leq 0.001$ compared to AngII in a Kruskal-Wallis multiple comparison test followed by a Dunn's post-hoc test.

HEK293 cells expressing the $AT_1R$ were assayed as described in the methods. $EC_{50}$ and $E_{max}$ are expressed as the means ± SD of values obtained in at least 3 independent experiments performed in triplicate.

**Table 3**

**Transduction ratios of AT$_1$R ligands**

| | AngII | | | | [Sar$^1$Hcy$^{3,5}$]AngII | | | | [Sar$^1$Cys$^3$Hcy$^5$]AngII | | | | [Sar$^1$Cys$^{3,5}$]AngII | | | |
|---|---|---|---|---|---|---|---|---|---|---|---|---|---|---|---|---|
| | $\log(\tau/K_A)$ | | $\Delta\log(\tau/K_A)$ | | $\log(\tau/K_A)$ | | $\Delta\log(\tau/K_A)$ | | $\log(\tau/K_A)$ | | $\Delta\log(\tau/K_A)$ | | $\log(\tau/K_A)$ | | $\Delta\log(\tau/K_A)$ | |
| G$_{q/11}$ | 8.49 | ± 0.11 | 0.00 | ± 0.16 | 6.79 | ± 0.08 | -1.70 | ± 0.14 | 5.45 | ± 0.11 | -3.05 | ± 0.16 | 4.76 | ± 0.19 | -3.74 | ± 0.22 |
| G$_{i2}$ | 8.84 | ± 0.33 | 0.00 | ± 0.47 | 8.36 | ± 0.26 | -0.48 | ± 0.42 | 5.87 | ± 0.19 | -2.97 | ± 0.39 | 5.30 | ± 0.25 | -3.54 | ± 0.42 |
| G$_{i3}$ | 8.97 | ± 0.28 | 0.00 | ± 0.39 | 8.98 | ± 0.13 | 0.01 | ± 0.30 | 7.18 | ± 0.12 | -1.78 | ± 0.30 | 6.62 | ± 0.20 | -2.35 | ± 0.34 |
| G$_z$ | 8.34 | ± 0.23 | 0.00 | ± 0.32 | 8.29 | ± 0.10 | -0.05 | ± 0.25 | 6.34 | ± 0.14 | -2.00 | ± 0.27 | 6.46 | ± 0.22 | -1.88 | ± 0.32 |
| βarr1 | 8.39 | ± 0.07 | 0.00 | ± 0.10 | 8.25 | ± 0.08 | -0.14 | ± 0.11 | 5.14 | ± 0.22 | -3.25 | ± 0.23 | 5.08 | ± 0.11 | -3.31 | ± 0.13 |
| βarr2 | 8.39 | ± 0.09 | 0.00 | ± 0.12 | 8.33 | ± 0.04 | -0.06 | ± 0.10 | 6.15 | ± 0.09 | -2.24 | ± 0.12 | 5.28 | ± 0.10 | -3.11 | ± 0.14 |
| $G_{12}$ | 8.14 | ± | 0.00 | ± | 8.14 | ± | 0.01 | ± | 4.84 | ± | -3.30 | ± | 5.21 | ± | -2.92 | ± |
| | 0.17 | | 0.25 | | 0.14 | | 0.22 | | 0.26 | | 0.31 | | 0.21 | | 0.28 | |
| PKC-ERK | 8.32 | ± | 0.00 | ± | 7.92 | ± | -0.40 | ± | 6.27 | ± | -2.10 | ± | 5.63 | ± | -2.70 | ± |
| | 0.08 | | 0.12 | | 0.11 | | 0.14 | | 0.21 | | 0.22 | | 0.09 | | 0.12 | |
| EGFR-ERK | 8.59 | ± | 0.00 | ± | 7.97 | ± | -0.62 | ± | 6.22 | ± | -2.37 | ± | 5.80 | ± | -2.79 | ± |
| | 0.03 | | 0.05 | | 0.09 | | 0.09 | | 0.05 | | 0.06 | | 0.45 | | 0.45 | |

HEK293 cells expressing the $AT_1R$ were stimulated with the different analogs and responses were measured for 9 distinct signaling pathways. Data were analysed by nonlinear regression using the Operational Model equation as described in the methods to determine $\log(\tau/K_A)$. $\Delta\log(\tau/K_A)$ were calculated from $\log(\tau/K_A)$ using AngII as the reference ligand. Data are the mean ± SEM of 3-6 independent experiments performed in triplicate.

**Table 4 - Bias factors of AT$_1$R ligands - G proteins vs βarrestin**

| | AngII | | [Sar$^1$Hcy$^{3,5}$]-AngII | | [Sar$^1$Cys$^3$Hcy$^5$]- | | [Sar$^1$Cys$^{3,5}$]-AngII | |
|---|---|---|---|---|---|---|---|---|
| | $\Delta\Delta\log(\tau/K_A)$ | BF | $\Delta\Delta\log(\tau/K_A)$ | BF | $\Delta\Delta\log(\tau/K_A)$ | BF | $\Delta\Delta\log(\tau/K_A)$ | BF |
| βarrestin1 /G$_{q/11}$ | 0.00 ± 0.19 | 1.00 | 1.56 ± | 36.10 | -0.21 ± 0.28 | 0.62 | 0.43 ± 0.26 | 2.67 |
| βarrestin2/G$_{q/11}$ | 0.00 ± 0.20 | 1.00 | 1.65 ± | 44.29 | 0.81 ± 0.20 | 6.42 | 0.63 ± 0.26 | 4.23 |
| G$_{12}$/G$_{q/11}$ | 0.00 ± 0.29 | 1.00 | 1.71 ± | 50.99 | -0.25 ± 0.35 | 0.56 | 0.81 ± 0.35 | 6.53 |
| G$_{i2}$/ G$_{q/11}$ | 0.00 ± 0.50 | 1.00 | 1.22 ± 0.45 | 16.47 | 0.08 ± 0.42 | 1.20 | 0.19 ± 0.47 | 1.56 |
| G$_{i3}$/ G$_{q/11}$ | 0.00 ± 0.39 | 1.00 | 1.71 ± 0.30 | 51.38 | 1.26 ± 0.30 | 18.3 | 1.39 ± 0.34 | 24.28 |
| G$_z$/ G$_{q/11}$ | 0.00 ± 0.36 | 1.00 | 1.65 ± 0.28 | 44.93 | 1.04 ± 0.31 | 11.0 | 1.85 ± 0.39 | 71.48 |
| βarrestin2/ G$_{12}$ | 0.00 ± 0.28 | 1.00 | -0.06 ± 0.24 | 0.87 | 1.06 ± 0.34 | 11.4 | -0.19 ± 0.31 | 0.65 |
| βarrestin1/ G$_{12}$ | 0.00 ± 0.27 | 1.00 | -0.15 ± 0.25 | 0.71 | 0.04 ± 0.39 | 1.10 | -0.39 ± 0.30 | 0.41 |
| βarrestin2/ βarrestin1 | 0.00 ± 0.16 | 1.00 | -0.09 ± 0.14 | 0.82 | -1.02 ± | 0.10 | -0.20 ± 0.19 | 0.63 |
| G$_{i2}$/ βarrestin1 | 0.00 ± 0.48 | 1.00 | -0.34 ± 0.44 | 0.46 | 0.29 ± 0.45 | 1.93 | -0.24 ± 0.44 | 0.58 |
| G$_{i3}$/ βarrestin1 | 0.00 ± 0.41 | 1.00 | 0.15 ± 0.32 | 1.42 | 1.47 ± 0.38 | 29.6 | 0.96 ± 0.37 | 9.08 |
| G$_z$/ βarrestin1 | 0.00 ± 0.34 | 1.00 | 0.10 ± 0.27 | 1.24 | 1.25 ± 0.36 | 17.8 | 1.43 ± 0.34 | 26.73 |
| G$_{i2}$/ βarrestin2 | 0.00 ± 0.49 | 1.00 | -0.43 ± 0.43 | 0.37 | -0.73 ± 0.41 | 0.19 | -0.43 ± 0.44 | 0.37 |
| G$_{i3}$/ βarrestin2 | 0.00 ± 0.41 | 1.00 | 0.06 ± 0.32 | 1.16 | 0.46 ± 0.33 | 2.86 | 0.76 ± 0.37 | 5.73 |
| G$_z$/ βarrestin2 | 0.00 ± 0.35 | 1.00 | 0.01 ± 0.27 | 1.01 | 0.24 ± 0.30 | 1.72 | 1.23 ± 0.34 | 16.88 |
| G$_{i2}$/ G$_{12}$ | 0.00 ± 0.53 | 1.00 | -0.49 ± 0.48 | 0.32 | 0.33 ± 0.50 | 2.12 | -0.62 ± 0.50 | 0.37 |
| G$_{i3}$/ G$_{12}$ | 0.00 ± 0.46 | 1.00 | 0.00 ± 0.38 | 1.01 | 1.51 ± 0.44 | 32.6 | 0.57 ± 0.44 | 5.73 |
| G$_z$/ G$_{12}$ | 0.00 ± 0.41 | 1.00 | -0.05 ± 0.34 | 0.88 | 1.29 ± 0.41 | 19.6 | 1.04 ± 0.42 | 16.88 |
| G$_{i3}$/ G$_{i2}$ | 0.00 ± 0.61 | 1.00 | 0.49 ± 0.52 | 3.12 | 1.19 ± 0.49 | 15.3 | 1.19 ± 0.54 | 0.24 |
| G$_z$/ G$_{i2}$ | 0.00 ± 0.57 | 1.00 | 0.44 ± 0.49 | 0.37 | 0.97 ± 0.47 | 9.26 | 1.66 ± 0.52 | 3.72 |
| G$_{i3}$/ G$_z$ | 0.00 ± 0.51 | 1.00 | 0.06 ± 0.39 | 1.14 | 0.22 ± 0.41 | 1.66 | -0.47 ± 0.47 | 10.95 |

$\Delta\Delta\log(\tau/K_A)$ and BF values were calculated as described in the methods. Data are the mean ± SEM of 3-6 independent experiments performed in triplicate. ** $P < 0.01$ and *** $P < 0.001$ in a Two-tailed unpaired t-test.

**Table 5 - Bias factors of AT$_1$R ligands – ERK signaling pathways**

| | AngII | | [Sar$^1$Hcy$^{3,5}$]-AngII | | [Sar$^1$Cys$^3$Hcy$^5$]- | | [Sar$^1$Cys$^{3,5}$]-AngII | |
|---|---|---|---|---|---|---|---|---|
| | $\Delta\Delta\log(\tau/K_A)$ | BF | $\Delta\Delta\log(\tau/K_A)$ | BF | $\Delta\Delta\log(\tau/K_A)$ | BF | $\Delta\Delta\log(\tau/K_A)$ | BF |
| PKC-ERK/βarrestin2 | 0.00 ± 0.17 | 1.00 | -0.35 ± 0.17 | 0.45 | 0.18 ± 0.26 | 1.52 | 0.41 ± 0.18 | 2.60 |
| PKC-ERK/βarrestin1 | 0.00 ± 0.15 | 1.00 | -0.26 ± 0.17 | 0.55 | 1.20 ± 0.32** | 15.7 | 0.61 ± 0.18 | 4.11 |
| EGFR-ERK/βarrestin2 | 0.00 ± 0.13 | 1.00 | -0.56 ± 0.13 | 0.27 | -0.13 ± 0.14 | 0.74 | 0.32 ± 0.47 | 2.10 |
| EGFR-ERK/βarrestin1 | 0.00 ± 0.11 | 1.00 | -0.47 ± 0.14 | 0.34 | 0.88 ± 0.24** | 7.64 | 0.52 ± 0.47 | 3.33 |
| PKC-ERK/G$_{12}$ | 0.00 ± 0.27 | 1.00 | -0.41 ± 0.26 | 0.39 | 1.24 ± 0.39 | 17.3 | 0.23 ± 0.30 | 1.68 |
| EGFR-ERK/G$_{12}$ | 0.00 ± 0.25 | 1.00 | -0.62 ± 0.24 | 0.24 | 0.93 ± 0.32 | 8.42 | 0.13 ± 0.53 | 1.36 |
| PKC-ERK/EGFR-ERK | 0.00 ± 0.12 | 1.00 | 0.21 ± 0.17 | 1.64 | 0.31 ± 0.23 | 2.06 | 0.09 ± 0.47 | 1.24 |
| PKC-ERK/G$_{q/11}$ | 0.00 ± 0.19 | 1.00 | 1.30 ± | 19.90 | 0.99 ± 0.27 | 9.78 | 1.04 ± | 10.99 |
| EGFR-ERK/G$_{q/11}$ | 0.00 ± 0.16 | 1.00 | 1.08 ± | 12.13 | 0.68 ± 0.17 | 4.74 | 0.95 ± 0.51 | 8.90 |
| G$_{i2}$/ PKC-ERK | 0.00 ± 0.49 | 1.00 | -0.08 ± 0.45 | 0.83 | -0.91 ± 0.45 | 0.12 | -0.85 ± 0.44 | 0.14 |
| G$_{i3}$/ PKC-ERK | 0.00 ± 0.41 | 1.00 | 0.41 ± 0.33 | 2.58 | 0.27 ± 0.38 | 1.88 | 0.34 ± 0.36 | 2.21 |
| G$_z$/ PKC-ERK | 0.00 ± 0.34 | 1.00 | 0.35 ± 0.28 | 2.26 | 0.05 ± 0.35 | 1.13 | 0.81 ± 0.34 | 6.51 |
| G$_{i2}$/ EGFR-ERK | 0.00 ± 0.48 | 1.00 | 0.13 ± 0.43 | 1.36 | -0.60 ± 0.39 | 0.25 | -0.76 ± 0.62 | 0.17 |
| G$_{i3}$/ EGFR-ERK | 0.00 ± 0.39 | 1.00 | 0.63 ± 0.32 | 4.24 | 0.59 ± 0.31 | 3.87 | 0.44 ± 0.57 | 2.73 |
| G$_z$/ EGFR-ERK | 0.00 ± 0.33 | 1.00 | 0.57 ± 0.27 | 3.70 | 0.37 ± 0.28 | 2.34 | 0.90 ± 0.55 | 8.04 |

$\Delta\Delta\log(\tau/K_A)$ and BF values were calculated as described in the methods. Data are the mean ± SEM of 3-6 independent experiments performed in triplicate. ** $P < 0.01$ and *** $P < 0.001$ in a Two-tailed unpaired t-test.

**Table 6**

**Rank order of potency of AngII and macrocyclic analogues on affinity and signaling pathways**

| Affinity | AngII > [Sar$^1$Hcy$^{3,5}$]AngII > [Sar$^1$Cys$^3$Hcy$^5$]AngII > [Sar$^1$Cys$^{3,5}$]AngII |
|---|---|
| IP-One | AngII > [Sar$^1$Hcy$^{3,5}$]AngII > [Sar$^1$Cys$^3$Hcy$^5$]AngII > [Sar$^1$Cys$^{3,5}$]AngII |
| G$_{i2}$ | AngII = [Sar$^1$Hcy$^{3,5}$]AngII > [Sar$^1$Cys$^3$Hcy$^5$]AngII > [Sar$^1$Cys$^{3,5}$]AngII |
| G$_{i3}$ | AngII = [Sar$^1$Hcy$^{3,5}$]AngII > [Sar$^1$Cys$^3$Hcy$^5$]AngII > [Sar$^1$Cys$^{3,5}$]AngII |
| G$_z$ | AngII = [Sar$^1$Hcy$^{3,5}$]AngII > [Sar$^1$Cys$^3$Hcy$^5$]AngII = [Sar$^1$Cys$^{3,5}$]AngII |
| βarr1 | AngII = [Sar$^1$Hcy$^{3,5}$]AngII > [Sar$^1$Cys$^3$Hcy$^5$]AngII = [Sar$^1$Cys$^{3,5}$]AngII |
| βarr2 | AngII = [Sar$^1$Hcy$^{3,5}$]AngII > [Sar$^1$Cys$^3$Hcy$^5$]AngII > [Sar$^1$Cys$^{3,5}$]AngII |
| G$_{12}$ | AngII = [Sar$^1$Hcy$^{3,5}$]AngII > **[Sar$^1$Cys$^{3,5}$]AngII > [Sar$^1$Cys$^3$Hcy$^5$]AngII** |
| PKC-ERK | AngII > [Sar$^1$Hcy$^{3,5}$]AngII > [Sar$^1$Cys$^3$Hcy$^5$]AngII > [Sar$^1$Cys$^{3,5}$]AngII |
| EGFR-ERK | AngII > [Sar$^1$Hcy$^{3,5}$]AngII > [Sar$^1$Cys$^3$Hcy$^5$]AngII > [Sar$^1$Cys$^{3,5}$]AngII |

Rank order of potency was determined based on Ki values for affinity and on the log($\tau$/K$_A$) for the signaling pathways.